\documentclass[11pt,twoside]{article}

\usepackage{asp2014}

\aspSuppressVolSlug
\resetcounters

\begin{document}

\title{Integrating UX Design in Astronomical Software Development: A Case Study}

\author{Yan~G.~Grange$^1$, and  Kevin~Tai $^1$}
\affil{$^1$ASTRON, Oude Hoogeveensedijk 4, 7991~PD, The Netherlands; \email{grange@astron.nl}; \email{tai@astron.nl}}

\paperauthor{Yan~G.~Grange}{grange@astron.nl}{0000-0001-5125-9539}{ASTRON}{Innovation and Systems}{Dwingeloo}{}{7991PD}{The Netherlands}
\paperauthor{Kevin~Tai}{tai@astron.nl}{}{ASTRON}{Innovation and Systems}{Dwingeloo}{}{7991PD}{The Netherlands}



\begin{abstract}
In 2023, ASTRON took the step of incorporating a dedicated User Experience (UX) designer into its software development process. This decision aimed to enhance the accessibility and usability of services providing access to the data holdings from the telescopes we are developing.

The field of astronomical software development has historically under emphasized UX design. ASTRON's initiative not only improves our own tools, but can also be used to demonstrate to the broader community the value of integrating UX expertise into development teams.

We discuss how we integrate the UX designer at the start of our software development lifecycle. We end with providing some considerations on how other projects could make use of UX knowledge in their development process.
\end{abstract}



\section{Introduction}
The concept of User Experience (commonly abbreviated as UX) can be defined the way users interact with a product, service or system. Within the field of UX, many techniques have been developed which aim to study users \textit{subjective} interactions with a system, to provide an \textit{objective} list of components that can be used to provide a \textit{uniform} experience, keeping in mind accessibility. Explicitly, this means that UX design should not be reduced to creating an aesthetically pleasing design. Even though users are probably happier interacting software that \textit{looks nice}, the main goal of UX design is to provide software that \textit{behaves predictably}. 
\par \citet{I301_adassxxxiii} provides a comprehensive overview of the applicability of UX thinking to astronomical software. Since 2023, ASTRON has had a dedicated UX designer working alongside and embedded in the software development process. In this paper, we describe how we have applied UX principles to develop some specific applications --- a proposal 
tool and a data access interface --- to improve the experience of the tools we build. These applications are part of a wider service ecosystem through which users can gain access to a telescope and its data. 

\section{Setting up a UX vision}
Creating a UX vision which describes the interactions users will have with the system is a way to support uniform and smooth access to the software and an important input to the requirements.
\articlefigure{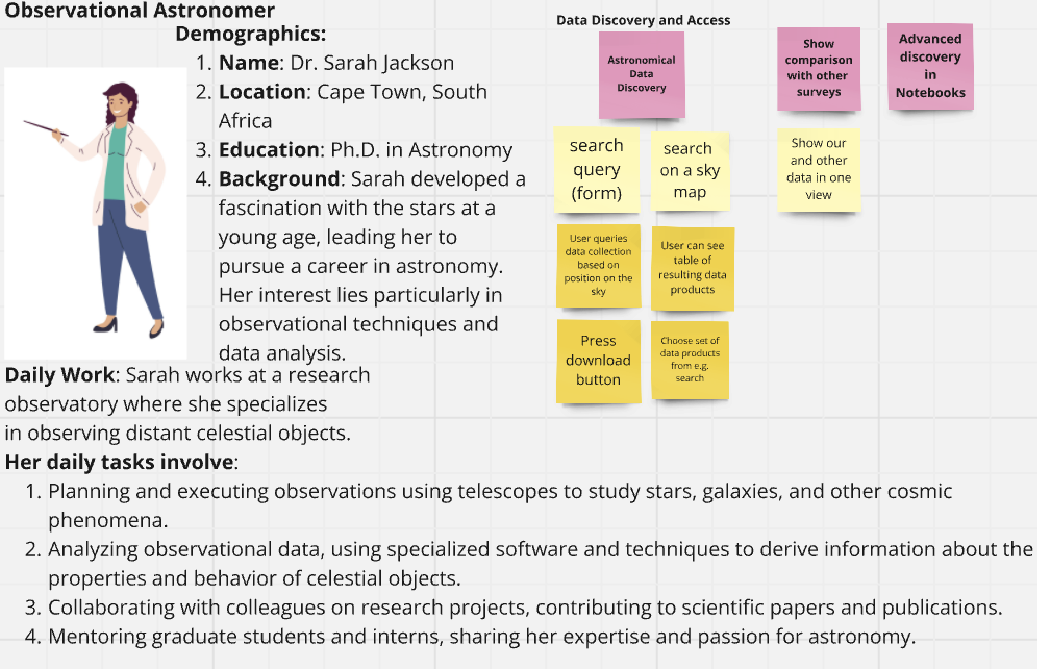}{fig:persona}{\textbf{Outside the dashed area}: Example of a user persona we define, in this case an optical astronomer named \textit{Sarah Jackson}. \textbf{Inside the dashed area}: Except from the analysis exercise for the \textit{Data Discovery and Access} function of our system, based on \textit{Sarah}s needs. From top to bottom, the stickies of different colour represent different granularity of components, if appropriate or needed. }

\par For defining the UX vision on the main system to access and analyse data products from LOFAR \citep{vanHaarlem2013}, we created a group of stakeholders consisting of our UX designer, the product owner\footnote{The role of the product owner is to ensure the development team is focused on work that delivers value}, the programme manager and user representatives. The first question this team should answer is \textit{who} the users are. For this we used the tool of defining a \textit{user persona}s. The goal of a persona is to define a \textit{type} of user, including relevant characteristics, as well as why they want to interact with the system. Giving a persona a name makes it easier to reason about a user (e.g. ``Sarah likes to query for RA and dec, but Fubuki just wants to enter a name''). To keep the discussion generic, it is however advised not to use existing users as personas. By defining personas that need to access a wide range of functionality, a user's flow through the system can be defined (also known as a \textit{user journey map}). For instance, one could put stickies on a (virtual) board defining each step in the process, as well as the functionality needed to perform this step and the elements that support this functionality. An example of a persona, and the outcome of the analysis of a few steps in the user flow is shown in Fig. \ref{fig:persona}. Then, the exercise can be repeated for the other personas, adding the bits of functionalities specific users need. 
\articlefigure{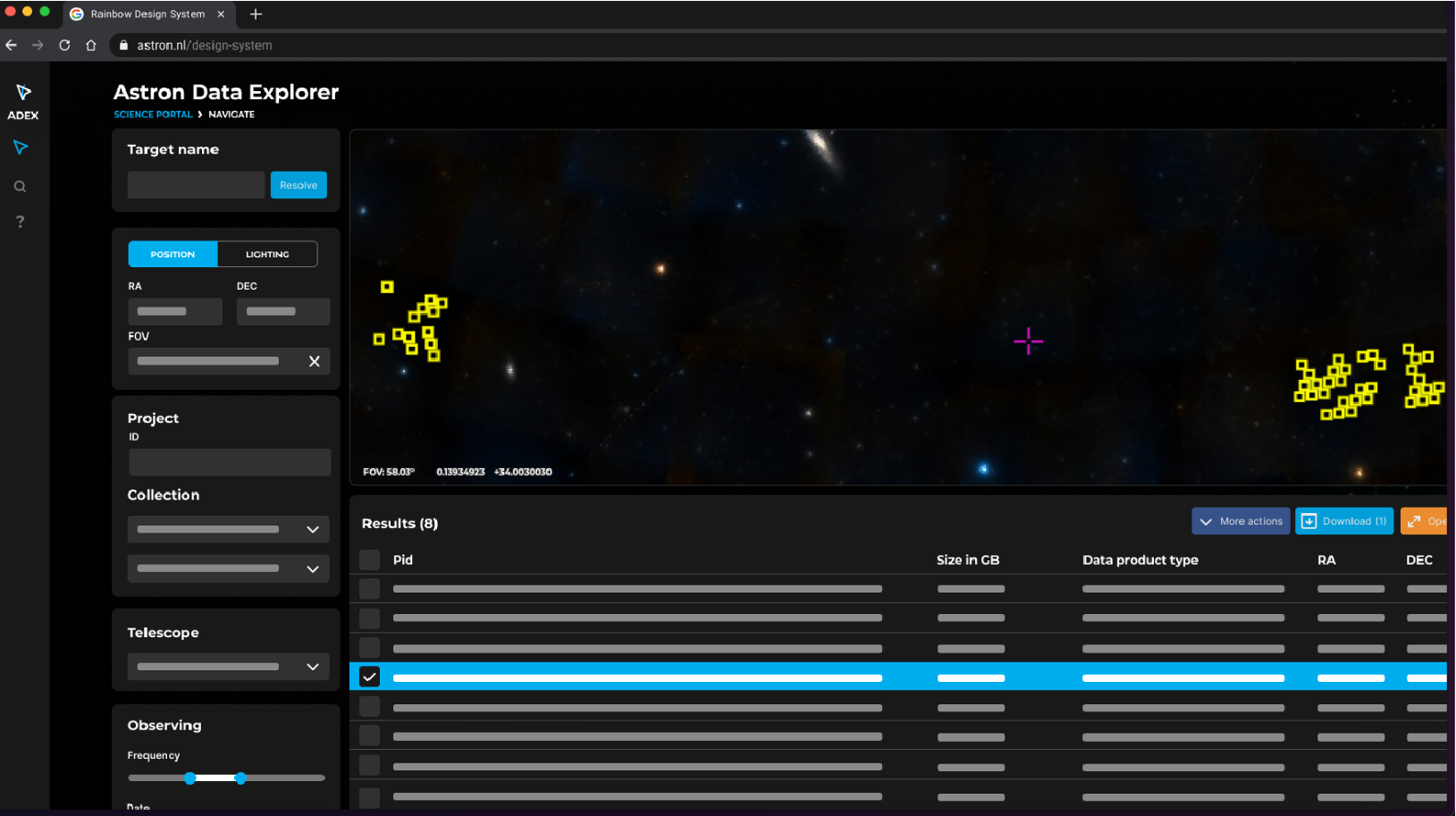}{fig:mockup}{An example of a mockup for a data query interface. By making the mockup clickable, users can be asked for feedback on the behaviour of the application before implementing any logic.}
\par Using the analysis the next step is to then create a (clickable) mockup\footnote{Several tools can be used to do this, like Figma or Marvel}, which acts as a quick prototyping of the typical way the persona would move through the system. This can then be discussed in the group, or with other groups of representative users and adapted based on the feedback. After several iterations, the path through the system is then finalised and the mockups can be used by the software developers to produce the actual application. An example of a mockup for the data query function can be seen in Fig. \ref{fig:mockup}.

\section{Design system}
\articlefigure{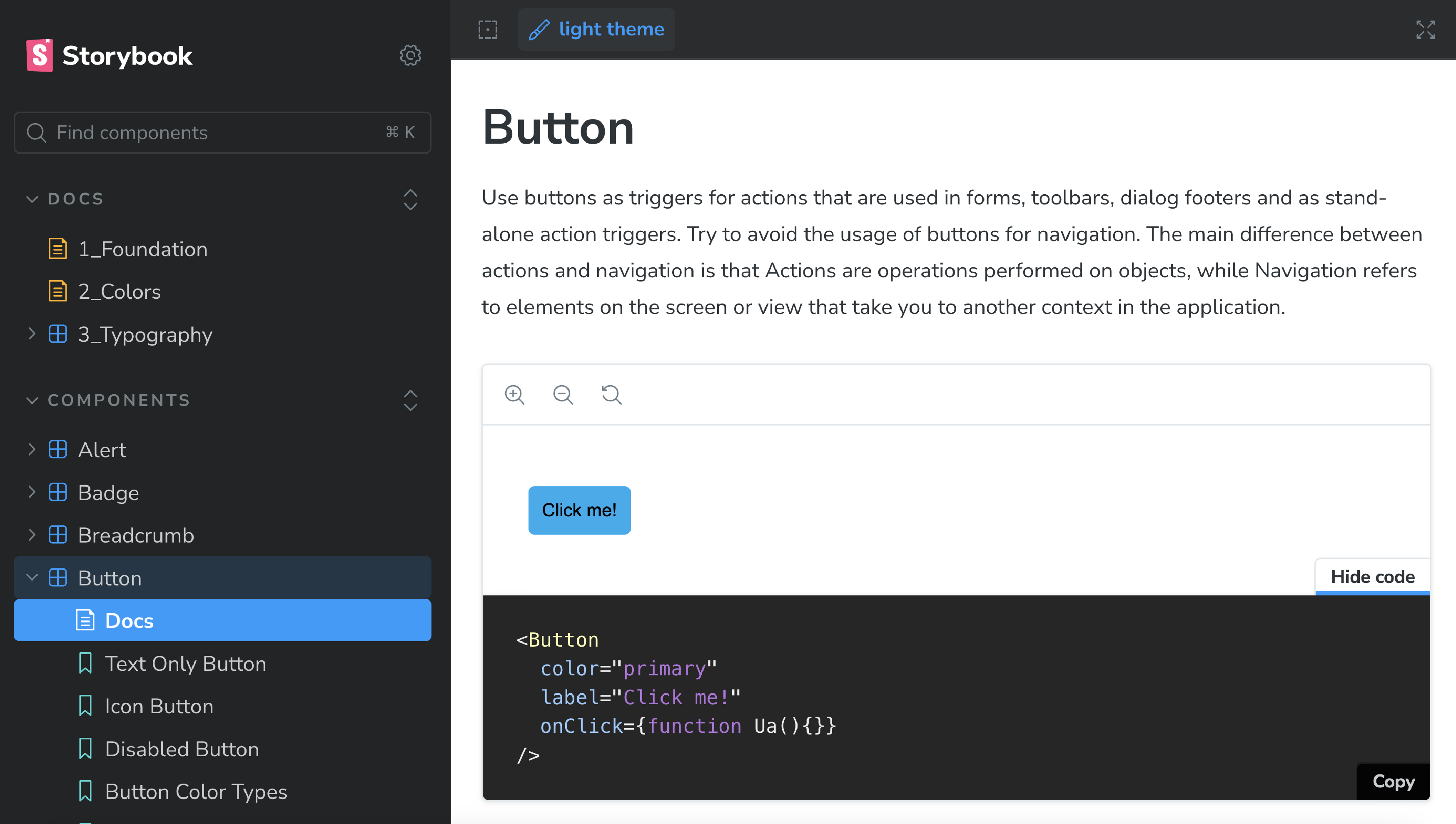}{fig:design_system}{View on the design system used in our development process. In this figure, the Button object is selected. The first element, called Docs, describes the type of object, shows how it looks and provides the user with all options that can be chosen (e.g. colour, icon, function), which then shows the actual code that can be used in the application. Below it in the menu, different types of button with specific properties can be shown (e.g. a submit button, or a button with an icon).}
When the vision has been translated to a prototype, we take the components (e.g. buttons, fields, date pickers), and put them in a central system (the \textit{design system}). A design system is a system where the choices and properties of a UX vision and corresponding design can be found. It contains general documentation, like foundational rationale, fonts, or colours, as well as versions of the components that can be reused by the developers when building the logic. The design system that we use at the ASTRON SDC \footnote{The design system used for the ASTRON SDC is publicly accessible and can be found here \url{https://sdc.astron.nl/design-system/}}, shown in Fig. \ref{fig:design_system}, makes it possible to copy the HTML code corresponding to a specific component so that it is guaranteed that components look and behave the same over all our tools, making the experience for users much smoother and predictable.
\par When the application has been built, based on the UX vision, and using the components from the design system, the development team will ask user representatives (which may have been involved with the vision group, but could also be different users to prevent biases) for feedback. In an iterative fashion the application (as well as the UX components) can then be improved.
\section{Considerations}
The initial effort of setting up a vision, creating personas and setting up a design system takes time. When the infrastructure is there, the amount of work needed will become less. Also the repository of components saves developers work when implementing the application and ensures the different applications to behave uniformly.
\par In this paper we have focussed on web-facing graphical user interfaces. However the techniques presented may very well be used in different cases as well. For instance, a mockup of a command line tool could uncover that the way the command line parameters are not as intuitive or logical as initially thought, or that the naming of a REST API could improve clarity.
\par Within the context of ASTRON, where we develop several services that need to work in together in a smooth and uniform way, it is very relevant to have a dedicated UX designer. For smaller projects, that may not be able to hire a dedicated UX designer, it is still of added value to perform an analysis of user needs before building the actual application to uncover logical paths through the software product and provide uniformity.

\bibliography{C405}  


\end{document}